\documentclass{iaus}
\usepackage{graphicx}
\usepackage{url}

\begin{document}
\title[Gravitational Waves] %% give here short title %%
{Gravitational Waves\\and Time Domain Astronomy}

\author[Centrella, Nissanke, Williams]   %% give here short author list %%
{Joan Centrella$^1$, Samaya Nissanke$^2$ \and Roy Williams$^2$}

\affiliation{$^1$NASA Goddard Spaceflight Center, Greenbelt, MD USA \\
email: {\tt Joan.Centrella@nasa.gov} \\[\affilskip]
$^2$California Institute of Technology, Pasadena CA USA \\
email: {\tt samaya@tapir.caltech.edu, roy@caltech.edu}}

\pubyear{2012}
\volume{285}  %% insert here IAU Symposium No.
\pagerange{1--86}
% \date{?? and in revised form ??}
\setcounter{page}{1}
\jname{New Horizons in Time Domain Astronomy}
\editors{E. Griffin \& R. Hanisch, eds.}

\maketitle

\begin{abstract}
The gravitational wave window onto the universe will open in roughly five years, when Advanced LIGO and 
Virgo achieve the first detections of high frequency gravitational waves, most likely coming from compact 
binary mergers.  Electromagnetic follow-up of these triggers, using radio, optical, and high energy telescopes, 
promises exciting opportunities in multi-messenger time domain astronomy.  In the decade, space-based 
observations of low frequency gravitational waves from massive black hole mergers, and their electromagnetic 
counterparts, will open up further vistas for discovery.   This two-part workshop featured brief presentations 
and stimulating discussions on the challenges and opportunities presented by gravitational wave astronomy.  
Highlights from the workshop, with the emphasis on strategies for electromagnetic follow-up, are presented in 
this report.
\end{abstract}

\section{New Cosmic Messengers}
Gravitational waves (GWs) are a new type of cosmic messenger, bringing direct information about the 
properties and dynamics of sources such as compact object mergers and stellar collapse.  
The observable GW spectrum spans over 18 orders of magnitude in frequency, 
ranging from phenomena generated in the earliest moments of 
the Universe to vibrations of stellar-mass black holes (BHs).

For time domain astronomy, two GW frequency bands stand out as being especially promising. The high 
frequency band,  $\sim$($1 - 10^4$) Hz, will be opened by ground-based interferometric detectors starting around 
mid-decade.  The strongest sources in this band are expected to be the mergers of stellar-mass compact objects, 
primarily BH binaries, neutron star (NS) binaries, and BH-NS binaries in the local universe out to  $\sim$several 
hundred Mpc.  The low frequency band,  $\sim$($10^{-4} - 10^{-1}$) Hz, 
will be opened by space-based detectors in the 2020s.
Merging massive black hole (MBH) binaries, with masses in the range $\sim$($10^3 - 10^7$) $M_{\odot}$,
detectable out to high redshifts $z > 10$, are expected to be the strongest sources here and the ones of greatest 
interest to time domain astronomy.  
Other sources include inspirals of compact objects into central MBHs in 
galaxies and compact stellar binaries with periods of tens of minutes
to hours\footnote{The very low frequency band, 
$\sim$($10^{-9} - 10^{-6}$) Hz, will be opened by pulsar timing arrays later this decade.  
The most likely sources in this region are binaries containing MBHs of $\sim10^9 M_{\odot}$.}.

The GWs emitted by binaries typically evolve upwards in frequency with time. When the binary 
components are well-separated and spiraling together due to GW emission, the waveform is a sinusoid 
increasing in frequency and amplitude, also called a chirp.  The final merger, in which the binary components 
coalesce, produces a burst of radiation; this is followed by a ringdown phase in which the merged remnant 
typically settles down to an equilibrium configuration.  The timescales for these phases depend primarily on the 
masses of the binary components.  For compact remnants, the inspiral might be detectable for about a 
minute, while the merger and ringdown occur on timescales of roughly a few to tens of milliseconds. The 
detailed nature of the merging objects, for example moduli of the neutron star crust, is revealed in these last 
few milliseconds. For MBH binaries, space-based detectors typically will be able to observe the inspiral for 
several months.  The ensuing merger and ringdown then occur over minutes to hours. 

The possibility of electromagnetic (EM) counterparts of these GW sources raises exciting prospects for multi-
messenger astronomy in the time domain. An EM counterpart greatly boosts confidence in a GW 
detection [\cite{kp93}]. EM counterparts may be a precursor, possibly associated with the 
binary inspiral; a flash, triggered during the merger and/or ringdown phases; or an afterglow. The timescales for 
these signals depend on emissions from gas in the vicinity of the binary and can vary widely depending on the 
type of EM emission produced.  Afterglows in particular can be very long-lived.

The opportunities for GW-EM multi-messenger time domain astronomy generated lively and fruitful discussions 
during a two-session workshop at the symposium.  This report presents the highlights from the workshop.

\section{High Frequency Gravitational Waves:  Getting Ready for Detection}
The GW window onto the universe is expected to open around the middle of this decade, when the GW signals 
from compact binaries are first detected by ground-based interferometers with kilometer-scale arms. These first 
detections will mark the culmination of decades of development by hundreds of scientists world-wide, and will 
inaugurate a new era in GW astronomy. This section begins with a status report on these efforts, and then 
highlights important questions and challenges for this new era in multi-messenger time domain astronomy.

\subsection{Status Update}
Currently there are three full-scale ground-based interferometric observatories: LIGO runs the
4-km observtories located in Hanford, Washington and Livingston, Louisiana, and the 3-km French-Italian 
Virgo detector
located near Pisa, Italy.  Both LIGO and Virgo were planned to be developed in stages. The initial 
detectors would be full-scale interferometers able to detect rare (nearby) events; the advanced detectors would 
be about a factor of 10 more sensitive, able to make multiple detections per year and be true observational 
tools.

LIGO and Virgo have successfully reached their initial design sensitivities, 
completing several science data-taking runs in 2009-2010. By current astrophysical estimates of stellar-mass 
mergers, this period of months of observation would have yielded a detection with probability $<2\%$ [\cite{abadie2010}]. 
Rather than waiting decades for a strong enough signal, LIGO and Virgo are undergoing upgrades to make 
joint detection rates at least yearly, perhaps weekly. Early science runs with Advanced LIGO/Virgo
could start by mid-decade.  
As of September 2011, Virgo is still operating; the upgrade to Advanced 
Virgo is expected to follow the Advanced LIGO upgrade by about a year.  Construction has also 
begun in Japan for the Large-Scale Cryogenic Gravitational-wave Telescope (LCGT, [\cite{kuroda2010}]), an advanced 
detector that could start operating by the end of this decade.  Another advanced detector may also be built in 
India.

The initial LIGO and Virgo detectors have been used to carry out science runs, deriving upper limits on sources 
within their observational reach, and to develop data-analysis and detection strategies.  An EM follow-up 
program and a blind-injection test were exercised during the 2009-2010 science runs, providing valuable 
experience for the advanced detector era, and having particular significance for this workshop.  
The follow-up program incorporated a prompt search for EM counterparts triggered by GW transients 
[\cite{Abbott:2011ys}].  Candidate GW events and their possible sky locations were identified using a low-latency 
analysis pipeline. The most promising sky positions for EM imaging were selected using a catalog of nearly 
galaxies and Milky Way globular clusters.   This directional information was sent to partner telescopes around 
the globe, as well as the Swift gamma ray satellite, (see Figure 1) within $\sim$30 min.  Nine such events were 
followed up by at least one telescope [\cite{Abbott:2011ys}].

\begin{figure}[h]
%\begin{center}
\centerline{\includegraphics[width=0.88\textwidth]{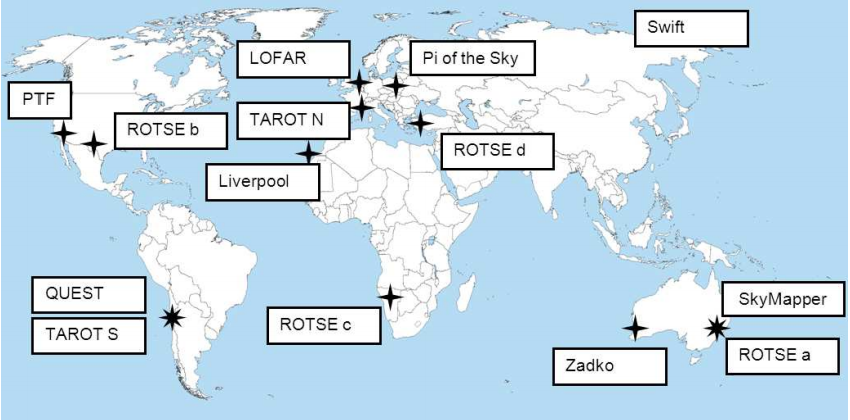}}
%\end{center}
\caption{Telescopes that participated in the LIGO-Virgo EM follow-up exercise.  Figure from Ref.
[\cite{Abbott:2011ys}]}
\end{figure}

A blind injection test was also carried out during September 2010 [\cite{BigDog}]. 
In this process, designed to provide 
a stress-test for the full data analysis pipeline and science procedures, signals were secretly injected into the 
detector data stream by a small group of analysts.  The parameters of these signals were sealed in an envelope, 
to be opened only after the full collaboration had searched the data and carried out a full exercise of the 
processes leading from potential detection to approval of a publication.  In this case, a strong chirp signal was 
observed in the data shortly after injection.  The data was vetted, sky positions were determined, and the GW 
trigger information was sent to several EM telescopes.   As part of the process, LIGO and Virgo prepared a data 
release representative of what could be released to the community when the first actual detections are made; 
see [\cite{BigDogData}].

\subsection{Expectations about GW Source Detection}
GWs are produced by the dynamical motion of massive objects in spacetime.  Since they couple very weakly to 
matter, GWs are ideal cosmic messengers to bring information from dark hidden regions within galaxies.  This 
section highlights some key expectations gleaned from studies of GW source detection, focusing on compact 
binary sources.

{\bf How many compact binaries are expected to be observed by advanced ground-based GW detectors?}
Rates for 
detection by Advanced LIGO and Virgo are estimated using projected
detector sensitivities, and compact binary merger rates derived from either the observed sample of galactic binary pulsars or population
synthesis results [\cite{abadie2010}]. A network comprising three advanced detectors are 
expected to detect between 0.4 and 400 NS/NS binaries per year, with the most realistic estimate being about 40 
per year, out to a distance of $\sim$450 Mpc for optimally oriented sources (that is, face-on and located 
directly above the detector).  For stellar BH binaries (with each BH having mass $\sim10 M_{\odot}$), the 
rates range from 0.2 to 300 detections per year, most realistically $\sim$20 per year, out to $\sim$2000 Mpc.  And for 
NS-BH binaries (where the BH has a mass $\sim10 M_{\odot}$), the rates range from 0.2 to 300 per year, with the most realistic estimate being $\sim$10 per year, 
out to $\sim$900 Mpc.  Averaging over the sky and the source orientation reduces these all distance ranges by a 
factor of about two [\cite{Finn}].

{\bf What properties of these sources can be measured by observing the GWs emitted?}
GWs carry direct 
information about their sources.  Broadly speaking, applying parameter estimation techniques to analyze the 
gravitational waveforms yields the binary masses, spins, and orbital elements, as well as extrinsic parameters 
such as the distance and position on the sky [\cite{Cutler:1994ys}]. 
Eventually it is hoped that the global GW network 
can elucidate deep secrets of extreme matter, ripped by space itself, through analysis of the merger waveform.

{\bf How well can these sources be localized on the sky?}
GW detectors are all-sky monitors.  A single 
interferometer has a broad antenna pattern and poor directional sensitivity.  Comparing the time of arrival of 
the signals in more than one detector allows for localization of the source on the sky.  Using the network of 
three advanced detectors (Virgo plus the two LIGO sites), the source's sky position can be reconstructed to a 
region of $\sim$1 to 100 deg$^2$ [e.g. \cite{Fairhurst, Wen, Nissanke:2011ax}].   
Adding additional interferometers, in particular one located out of 
the LIGO-Virgo plane such as LIGO India, brings improvements of
factors $\sim$3 [\cite{Schutz:2011}].  Note that the search regions are 
irregularly shaped and can have non-contiguous `islands' if the signals are near threshold (see Figure 2), as 
expected for many LIGO-Virgo sources.

\begin{figure}[h]
%\begin{center}
\centerline{\includegraphics[width=0.88\textwidth]{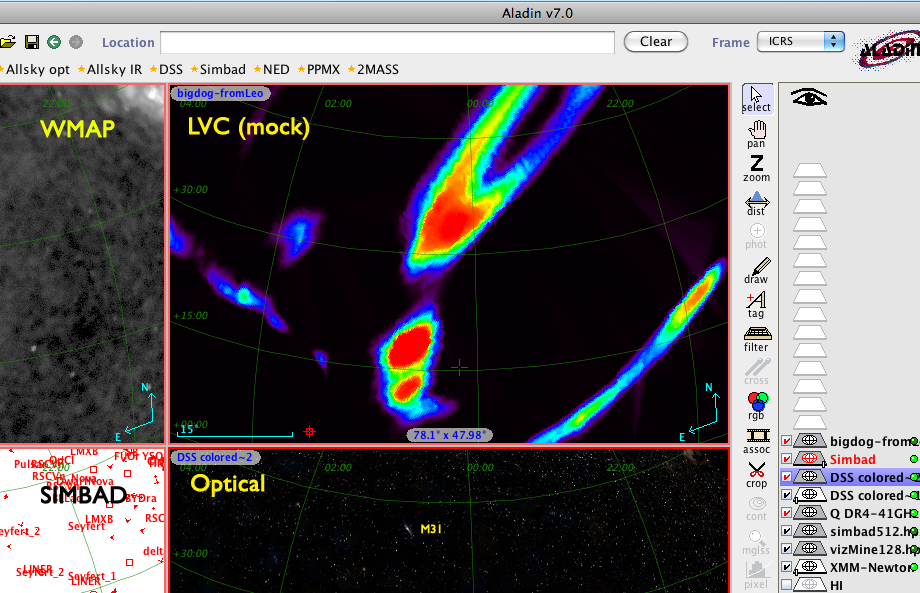}}
%\end{center}
\caption{Non-contiguous search islands, illustrated with the Aladin `astronomical information system' [\cite{Aladin}]. 
By converting the skymap to Healpix/FITS format,
it can be compared directly with other astronmical images and catalogs.
The panel `LVC (mock)' corresponds to the skymap released as [\cite{BigDogData}], 
with at least seven islands visible. 
Other panels show the same sky with other maps for comparison: 
left: WMAP data, 
lower-left: SIMBAD graduated catalog, and 
right: the Optical sky (DSS), where the galaxy M31 can be seen. }
\end{figure}

{\bf What types of information can EM astronomers expect to receive from GW observatories?}
The advanced 
detector network will be able to supply the signal time of arrival and sky localization of the source, along with an 
estimate of the false alarm rate (FAR).  For merging compact binaries, the masses and spins of the components 
along with the inclination and luminosity distance will also be available.  Other information may also be 
released; see [\cite{BigDogData}] and Figure 2 for an example. 

{\bf When will LIGO publically release rapid alerts for afterglow observers?}
Rapid release of triggers ($\sim$minutes) will begin under one of three criteria, according to the LIGO Data 
Management Plan [\cite{DMP}]: after a GW detection has been confirmed and the collaboration agrees to begin
the public program; or after a large volume of space-time has been searched by LIGO; or if the detectors
have been running for a long time. It is hoped that this release could be as early as 2016. It should be noted, 
however, that these criteria may be changed in the future and the release date brought forward.

\subsection{Observing EM Counterparts:  Challenges and Opportunities}
Mergers of NS/NS and NS/BH binaries are expected to generate EM radiation in several wavelength bands.  
Coupling observations of EM radiation and GWs from these sources opens up exciting new avenues for 
exploration.  Since many GW detections may be near threshold, identification of an EM counterpart would 
provide additional confirmation of the event.  Direct measurements of the binary properties through GW 
observations will allow testing and deeper understanding of the underlying astrophysical models currently 
inferred only from EM radiation.  This section captures exciting and challenging issues in this arena.

{\bf Which astrophysical phenomena generate promising EM counterparts to NS-NS and NS-BH mergers?}
These 
mergers are expected to produce collimated jets, observed near the axis as short-duration gamma-ray bursts 
(SGRBs).  Afterglows from interactions of a jet with gas around the burst can be observed in the optical (near the 
axis, on timescales of hours to days) and radio (isotropically, over 
weeks to years).  A `kilonova' may also be 
produced from radioactive decay of heavy elements synthesized in the
ejecta, yielding weak optical afterglows lasting 
several days [\cite{Li:1998}].  Note that SGRBs are rare within the 
distance observable by Advanced LIGO and Virgo.  Thus the 
isotropic emissions, particularly those from kilonovae, are likely the most promising observable EM 
counterparts. See [\cite{Metzger:2011bv}] for details.

{\bf What information will EM astronomers want from the GW observatories?}
Most astronomers will first want 
the time and sky location of the event to do the follow-up; information about the binary components and orbits 
will be valuable for understanding the underlying astrophysics.  A smaller group of scientists may also want the 
gravitational wave time series information.  The means of distributing this information is currently being worked 
out.

{\bf Which astronomical instruments will be especially useful in searching for EM counterparts?}
Satellites such as 
Swift and Fermi are needed to find GRBs, which we expect to be coincident with GW detections; but only if the 
axis of the inspiral system points to the Earth (`beaming'). 
Other $\gamma$/X satellite observatories, current and planned, include MAXI, 
SVOM, and AstroSat. Rapid follow-up by ground-based optical telescopes with wide fields of view 
(such as PTF-2, Pan-STARRS, LSST and others) and radio 
telescopes (LOFAR, ASKAP, EVLA etc) will search for the afterglows.

{\bf Which EM follow-up strategies will produce the best astrophysical results?}
In the simplest scenario, all follow-
up telescopes operate independently, pointing at the regions identified as having the greatest probability of 
containing the target.  Since the source error regions on the sky from GW events will be relatively large 
irregularly-shaped regions that may have non-contiguous islands, many EM counterparts, particularly afterglows 
of relatively short duration, will be missed in this approach.  Coordinated EM follow-up, in which many 
telescopes operate cooperatively to cover the source region, can dramatically increase the odds of imaging an 
EM counterpart [\cite{Singer}].

{\bf Would it be possible to observe the immediate optical `flash', 
that is expected within seconds of the merger time?}
There are all-sky optical monitors in operation and in planning. The `Pi Of The Sky' observatory can locate 
such a very fast transient in time, to within seconds [\cite{PiOfTheSky}] 
to fainter than magnitude 12, and the planned 
global monitoring system from Los Alamos and LCOGT [\cite{Wren}] will have the advantage of much higher duty 
cycle, due to its distributed nature and multiple telescopes at each site.

{\bf How early a notice of a GW trigger is useful, and possible?}
Since early notification increases not only the odds 
of successfully imaging the counterpart, but also the amount and types of information that will be gained, send 
the notices as early as possible.  For the recent follow-up program carried out by LIGO and Virgo, the notices 
were sent out within 30 min of detection [\cite{Abbott:2011ys}]; efforts are continuing to 
reduce this latency to 1 minute.  
Detecting the GW signal during the binary inspiral (before the merger) and releasing this information before the 
burst occurs is even more interesting and challenging; see [\cite{Cannon:2011vi}] for an early study.

{\bf Which triggers should be followed up? How low a significance is tolerable?}
Nearby strong GW sources should have a low false-alarm rate (FAR) and produce more robust data, 
including more accurate values for the binary parameters and sky 
location.  But signals near threshold might be more common.  The answers to these questions will depend on 
the actual detection rate.

{\bf What more information about the local universe is needed to be ready for the data from Advanced LIGO and Virgo?}
Two items in particular are needed:  a complete publicly-available
catalog of nearby galaxies [\cite{White,Kulkarni:2009}], and an 
inventory of known transients within the reach of Advanced LIGO and Virgo.
It should be noted that for faint transients, it may be that lack of such a catalog, and large numbers, 
may make it very difficult to pick out a GW afterglow, even if there
is a wide, deep, telescope available. By using spatial coincidence with a complete 
galaxy catalog for those events within 200 Mpc, the number of false-positve 
transients in a typical GW error box of $\sim$10 deg$^2$ reduces by three orders of magnitude [\cite{Kulkarni:2009}].

\section{Low Frequency Gravitational Waves: Looking to the Future}
The low frequency window contains a wealth of astrophysical sources.  
Due to noise from fluctuating gravity gradients in the Earth at frequencies below a few Hz, low frequency GWs 
can only be observed using space-based detectors.  The most highly developed proposal is the Laser 
Interferometer Space Antenna (LISA), which consists of three satellites orbiting the Earth in a triangular 
configuration with arm lengths $\sim5 \times 10^6$ km [\cite{Jennrich:2009ti}].
Detectors such as LISA will observe coalescing MBH binaries inspiraling over a 
period of several months, followed by the final merger and ringdown,
as well as inspirals of compact objects into central MBHs in galaxies
and compact stellar binaries with periods of minutes to hours.

{\bf What is the current status of low frequency GW detectors?}
Due to budgetary problems, ESA and NASA 
terminated their partnership to develop LISA in the spring of 2011 and both agencies are now looking at lower-
cost concepts.  On the ESA side, studies are underway for the New Gravitational-wave Observatory (NGO, also 
known informally as LISA-lite, EuLISA, and eLISA), which is similar to LISA but with shorter arm lengths 
$\sim1 \times 10^6$ km. In the US, NASA is also examining concepts for a Space Gravitational-wave Observatory (SGO).  

{\bf What can be learned from observing low frequency GWs from MBH mergers?}
For these space-based 
interferometers, MBH mergers [\cite{Sesana:2010wy}] can be observed over a period of several months at relatively 
high signal-to-noise, allowing precision measurements of the binary properties, plus sky 
localization to $<100$ deg$^2$.   
The expected merger time can be predicted and broadcast weeks or months in advance, providing 
excellent opportunities for EM follow-up.  The rates for MBH mergers are expected to be at least several per 
year, with the actual values depending on the instrument sensitivity. 

{\bf What are the prospects for EM counterparts of MBH mergers?}
MBH mergers are astrophysically rich systems, 
with a variety of possible EM signals as precursors, flashes, and afterglows [\cite{Schnittman:2010wy}].  
Since MBH mergers are considered central to our understanding of galaxy and MBH assembly history and 
demography, and 
galaxy-MBH co-evolution, the astrophysical payoffs will be significant [\cite{Komossa:2003wz}].  

\section{Summary}
The GW window onto the universe will open this decade, when Advanced LIGO and Virgo make the first 
detections of high frequency GW signals, expected to come from merging compact binaries.  Strategies for EM 
follow-up of the GW triggers are being designed and tested to search for radio, optical, and high energy 
counterparts. Searches for coincident GW and high-energy neutrinos will be done in coordination with the 
IceCube project [\cite{HEN}]. Coordinated searches, coupled with complete catalogs of galaxies and transients in the 
local universe, are needed to maximize the science from these multi-messenger studies.  In the next decade, 
space-based interferometers will open the low frequency window with observations of merging MBH binaries. 
These observations, and their EM counterparts, will provide important information on the evolution of structure 
and MBHs over cosmic time.
 
\begin{acknowledgements}
It is a pleasure to thank Eric Chassande-Mottin, Ed Daw, Stephen Fairhurst, Jonathan Gair, Mansi Kasliwal, 
Stefanie Komossa, Brian Metzger, Larry Price, Jeremy Schnittman, Alberto Sesana, and Leo 
Singer, who delivered excellent short presentations at the GW workshops that stimulated interesting and 
important discussions.  
\end{acknowledgements}

\end{document}